\documentclass[5p,times]{elsarticle}

\usepackage{lineno,hyperref,graphicx}

\usepackage{gensymb}
\usepackage{amsmath}
\usepackage{caption}
\usepackage{subcaption}

\begin{document}

\begin{frontmatter}

\title{Stability studies of sealed Resistive Plate Chambers}
\author[lipAddress]{A. Blanco\corref{mycorrespondingasealeduthor}}
\cortext[mycorrespondingauthor]{Corresponding authors}
\ead{alberto@coimbra.lip.pt}
\author[lipAddress,isecAddress]{P. Fonte}
\author[lipAddress]{L. Lopes\corref{mycorrespondingauthor}}
\ead{luisalberto@coimbra.lip.pt}
\author[lipAddress,tecnicoAddress]{M. Pimenta}

\address[lipAddress]{Laborat\'orio de Instrumenta\c{c}\~ao e F\'isica Experimental de Part\'iculas (LIP), Departamento de F\'isica da Universidade de Coimbra, 3004-516 Coimbra, Portugal}
\address[isecAddress]{Instituto Polit\'ecnico de Coimbra, Instituto Superior de Engenharia de Coimbra, Rua Pedro Nunes, 3030-199 Coimbra, Portugal}
\address[tecnicoAddress]{Departamento de F\'isica, Instituto Superior T\'ecnico, Universidade de Lisboa Avenida Rovisco Pais, n. 1, 1049-001 Lisboa, Portugal}
\begin{abstract}
The phase-out of hydro-fluorocarbons, owing to their high Global Warming Power, affecting the main gas used in Resistive Plate Chambers (RPCs), tetrafluoroethane C$_2$H$_2$F$_4$, has increased operational difficulties on existing systems and imposes strong restrictions on its use in new systems. 

This has motivated a new line of R\&D on sealed RPCs: RPCs that do not require a continuous gas flow for their operation and dispense the use of very complex and expensive re-circulation and/or recycling gas systems. At the moment it is not clear whether this solution can cover all fields of application normally allocated to RPCs, but it seems that it could be considered as a valid option for low particle flux triggering/tracking of particles, e.g. in cosmic ray or rare event experiments.

In this work, we demonstrate the feasibility of a small telescope for atmospheric muon tracking consisting of four $300$~x~$300$~mm$^2$ sealed RPCs with gas gap widths of $1$~mm, $1.5$~mm and $2$~mm. The results suggest that it is possible to operate this type of detectors for extended periods of time (more than five months) with its main characteristics, efficiency, average charge and streamer probability, without apparent degradation  and similar to a RPC operated in continuous gas flow.
\end{abstract}


\end{frontmatter}

\section{Introduction}
Resistive Plate Chambers (RPC), like most gaseous detectors, rely on the purity of the gas to keep their performance stable. For this reason, they operate with re-circulation, purification and cleaning systems that attempt to keep the gas purity unchanged. Gas purity degradation is, in the first instance, primarily due to leaks/permeability in the system that allow atmospheric gases and/or humidity to enter, as well as the release of hydrofluorocarbons (HFCs), with C$_2$H$_2$F$_4$ as main constituent. This results in the need to inject substantial amounts of fresh gas. The use of HFCs, and in particular their release into the environment, is a serious problem today. This is due to the high global warming potential (GWP) of HFCs, which contributes significantly to global warming. In fact, the European Union (EU) decreed, already in $2015$, the phase-out of HFCs. This poses serious problems for existing RPC systems, but especially for new systems, which will certainly not be able to be planned in the same way. In addition, the gas system associated with these detectors introduces considerable complexity and cost.

A possible solution to the problem, from an environmental point of view (the problem of the complexity of the gas system would remain), is the replacement of these gases by others with a much lower GWP, the so-called green gases \cite{Man21}, with HFO-$1234$ze (C$_3$H$_2$F$_4$) as the most promising solution \cite{GARILLOT2024169104}. 

Another possible solution would be to construct and operate RPCs without any gas supply, i.e. RPCs that contain gas but are hermetically sealed after construction, similar to the popular Geiger-Mueller tube. This concept has been baptized as sealed RPCs (sRPC) \cite{Lopes_2020, Blanco2023}. It would mitigate the problem of HFCs phasing out by drastically minimizing (by orders of magnitude) the amount of gas used today, reducing its environmental impact to negligible levels. It would also eliminate any dependence on complex gas systems, allowing the expansion of this type of technology towards Cosmic Ray (CR) experiments through the construction of large, high-performance arrays at low cost, which would replace the simple Cerenkov water tanks in remote and difficult-to-access locations typical of CR experiments \cite{LOPES2023168446}.

In this work, we present the characterization and stability studies of four $300$~x~$300$~mm$^2$ sRPCs with gas gap widths of $1$~mm, $1.5$~mm and $2$~mm, by exposing them to the natural flux of cosmic rays.

\section{Experimental setup}\label{sec:setup}

\subsection{Sealed RPC modules}\label{sec:sealed}
\label{subsec:sealed}
\begin{figure}[h]%
	\centering
	\includegraphics[width=0.5\textwidth]{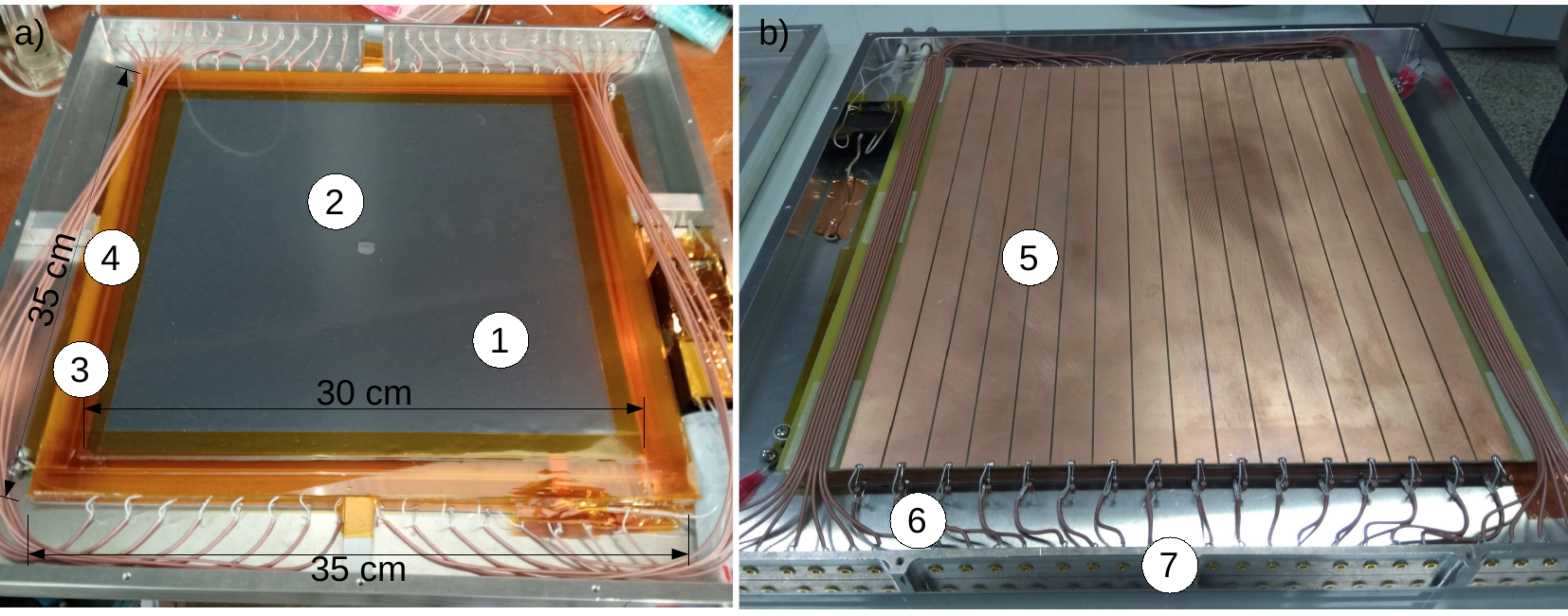}
	\caption{a) sRPC module showing: 1- HV layer, 2- Circular spacer in the center of the active area, 3- Strip spacer all around de the periphery and 4- Mylar\textsuperscript{TM} and  Kapton\textsuperscript{TM} layers. b) sRPC plane showing: 5- Readout strip plane, 6- Coaxial cables and  7-  MMCX RF feedthrough connectors.}\label{fig:setupSRPC}
\end{figure}

The sRPC modules are multi-gap structures \cite{CERRONZEBALLOS1996132} equipped with two gas gaps defined by three $2$~mm thick soda lime glass electrodes \footnote{with bulk resistivity of $\approx 5x10^{12}$~$\Omega$cm at $25$~$^\circ$C} of about $350$~x~$350$~mm$^2$ separated by spacers. Each gap includes a circular spacer made of a soda lime glass disk ($10$~mm diameter) placed in the center of the active area and a strip ($25$~mm width), also made of soda lime glass, all around the periphery. Three types of modules were constructed with different spacer thicknesses, defining different gas gaps, $1$~mm, $1.5$~mm and $2$~mm. For the assembly, the multi-gap structure is kept under controlled pressure and all peripheral surfaces are covered with an epoxy glue for sealing and mechanical strength. Gas inlets and outlets, one per gas gap, (used for gas filling) are made with standard plastic dispensing needles (from Loctite \textsuperscript{TM}). The High Voltage (HV) electrodes are made up of a semi-conductive\footnote{Based on an artistic acrylic paint with around $100~M\varOmega/\Box$.} layer airbrushed to the outer surface of the outermost glasses of the multi-gap structure in a area of $300$~x~$300$~mm$^2$ and covered with a Mylar\textsuperscript{TM} and Kapton\textsuperscript{TM} layers for electrical insulation, see figure \ref{fig:setupSRPC}.a for reference. After assembly the gas gaps are filled with a mixture of $97.5$\%  C$_{2}$H$_{2}$F$_{4}$ and $2.5$\% SF$_{6}$.

\subsection{Sealed RPC planes}
\label{subsec:readout}
The sRPC modules are read out by a readout strip plane\footnote{Made of $1.6$~mm thick Flame Retardant 4 (FR4) Printed Circuit Board (PCB).} equipped, in one side, with sixteen $18$~mm width, $19$~mm pitch and $40$~mm long copper strips, located on top of the module. A ground plane, placed on bottom completes the readout structure. The complete structure is enclosed in an aluminum box that provides the necessary electromagnetic insulation and mechanical rigidity. Every strip is connected at each end to a coaxial cable inner conductor, while the outer conductor is connected to the ground plane. The coaxial cables are connected to MMCX RF feedthrough connectors, see figure \ref{fig:setupSRPC}.b for reference. 
	
The thirty two sRPC plane signals are fed to fast Front End Electronics (fFEE) \cite{HADES_FEE} channels borrowed from the HADES RPC-TOF detector \cite{HADES_RPC} capable of measuring time and charge in a single channel. The resulting digital signals are read out by the TRB board \cite{TRB3} equipped with $128$ multi-hit TDC (TDC-in-FPGA technology) channels with a time precision better than $20$~ps. 

With this arrangement we are able to provide information on:
\begin{itemize}
	\item Charge, $Q$, as the sum of the induced charge on the strips (or the mean charge, $<Q>$, as the mean of the $Q$ distribution), and streamer probability as the percentage of events with a charge higher than $100$ (A.U.).
	\item Time, $T$, as the half-sum of the times at both ends for the strip with maximum charge.
	\item Longitudinal position to the strips, $Y$, as the difference of times at both ends for the strip with maximum charge multiplied by the propagation velocity on the strip, $\approx 165$ mm/ns.
	\item Transversal position to the strips, $X$, as the position of the strip with maximum charge. Both, longitudinal and transversal positions are determined with a resolution of about $10$ mm sigma. 
\end{itemize}
  
\subsection{sRPC telescope}
\begin{figure}[t]%
	\centering
	\includegraphics[width=0.5\textwidth]{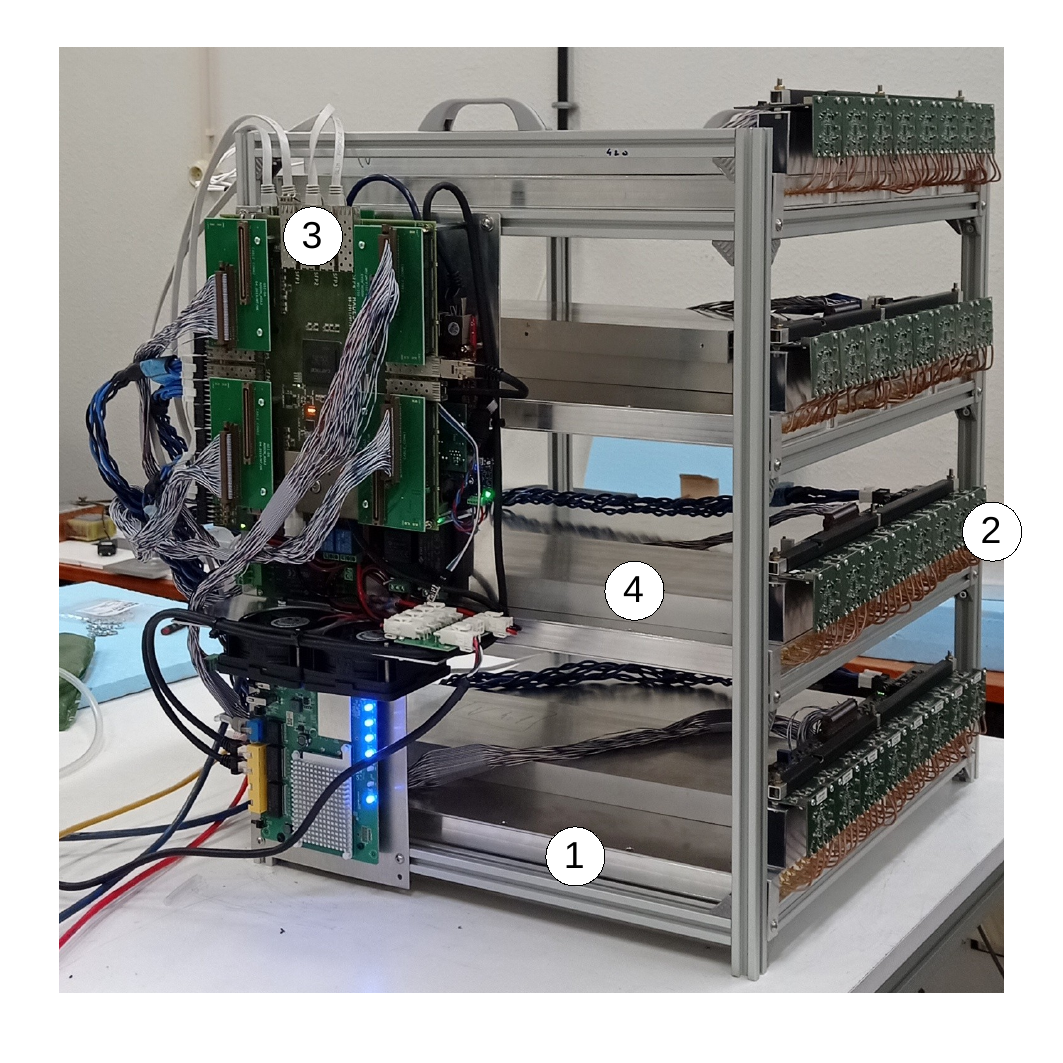}
	\caption{Experimental setup for sRPC characterization consisting of four detection planes equipped with sRPC modules of $300$~x~$300$~mm$^2$ vertically arranged at a distance of $20$~cm. 1- sRPC planes, 2- Fast Front End Electronics board housing 32 channels, 3- DAQ system and 4- HV power supply.}\label{fig:setup}
\end{figure}

The experimental setup for sRPC characterization consists of four detection planes equipped with sRPC modules, vertically arranged at a distance of $20$~cm, on a common structure, see figure \ref{fig:setup}, from now on called sRPC telescope or telescope.

The bottom and top planes, named RPC01 and RPC04 respectively, are equipped with $1$~mm gas gaps modules while the planes above bottom and below top, named RPC02 and RPC03, are equipped with the $2$ and $1.5$~mm gas gaps. Each detection plane is powered by a custom made HV power supply, allowing to control/monitor independently each of the planes, which is necessary due to the different gas gap widths present. The data acquisition (DAQ) system including: TRB board, low voltage power supplies and acquisition computer, is fixed on one side. This way, all the necessary instrumentation to operate the telescope is located in the same structure, forming a portable device.

With this setup, it is possible to characterize the response of each sRPC plane to atmospheric muons using the other three planes as trigger. Thus we define the efficiency as the inverse of the quotient of the number of events seen by three planes divided by the corresponding number of events seen by all four planes. As the telescope, has tracking capability it is possible to calculate the efficiency as a function of position, as well as other relevant quantities for the characterization of sRPCs.

\section{Results}

\begin{figure}
	\centering 
	\includegraphics[width=\linewidth]{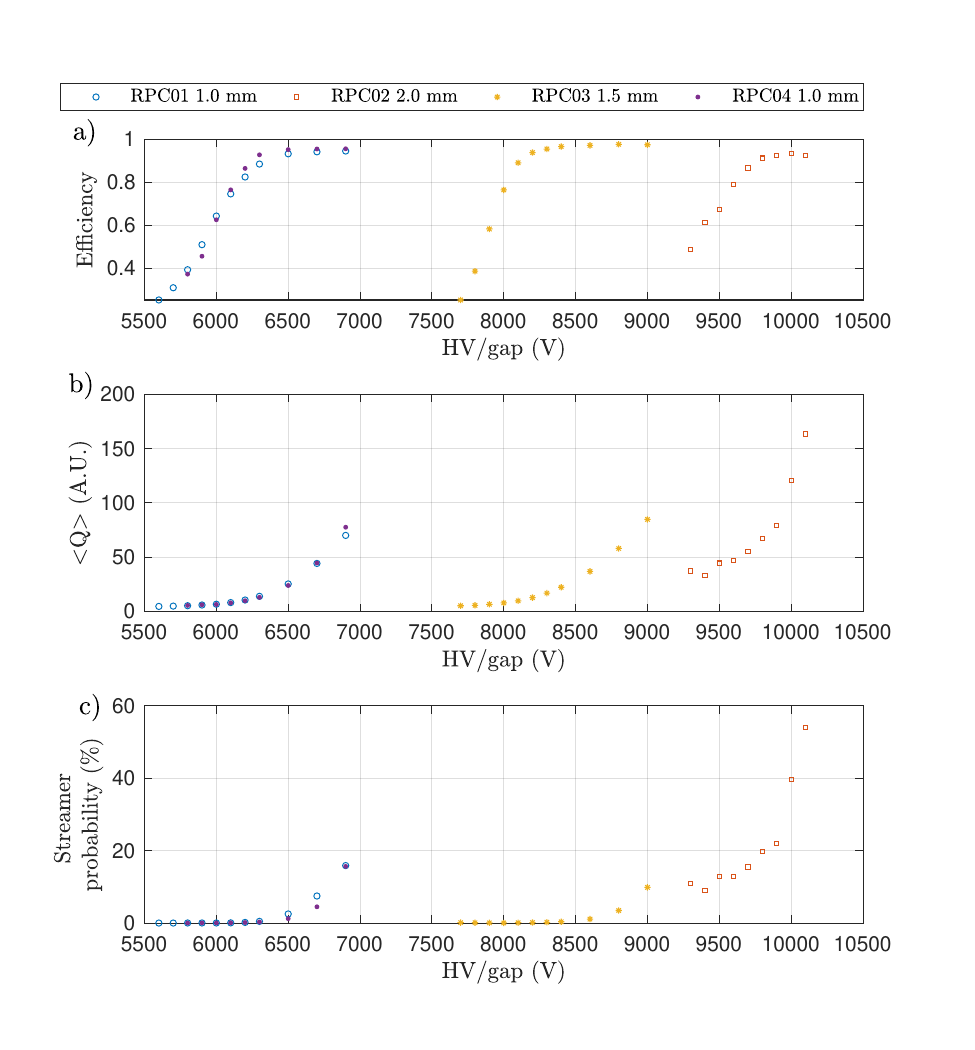}
	\caption{HV scan performed, progressively decreasing the HV of all planes, after having kept HV values within the HV plateau several months after sealing the sRPC modules. a) efficiency, b) average charge and c) streamer probability.}
	\label{fig:HVscan}
\end{figure}

Figure \ref{fig:HVscan} shows the results of an HV scan performed, progressively decreasing the HV of all planes, after having kept HV values within the HV plateau several months after sealing the sRPC. In general the behavior is identical to RPCs operated in a continuous gas flow. Figure \ref{fig:HVscan}.a plots the efficiency as a function of applied HV showing: HV plateaus with efficiency higher than $95 \%$ and about $500$~V wide (RPC02 has a somewhat lower value, $92$ \%, probably related to the high probability of streamers, see below) and the similitude between the $1$~mm gas gap planes. Figure \ref{fig:HVscan}.b  shows the average charge as a function of HV showing the similitude for all planes except for RPC02 which again has a different behavior related to the high probability of streamers. Finally, Figure \ref{fig:HVscan}.c shows the probability of streamers as a function of HV, very similar for all planes except for RPC02, which starts from values close to $10\%$ reaching up to about $50\%$. This behavior is attributed to a deficient initial sealing, corresponding to an older version technique.

\begin{figure}
	\centering 
	\includegraphics[width=\linewidth]{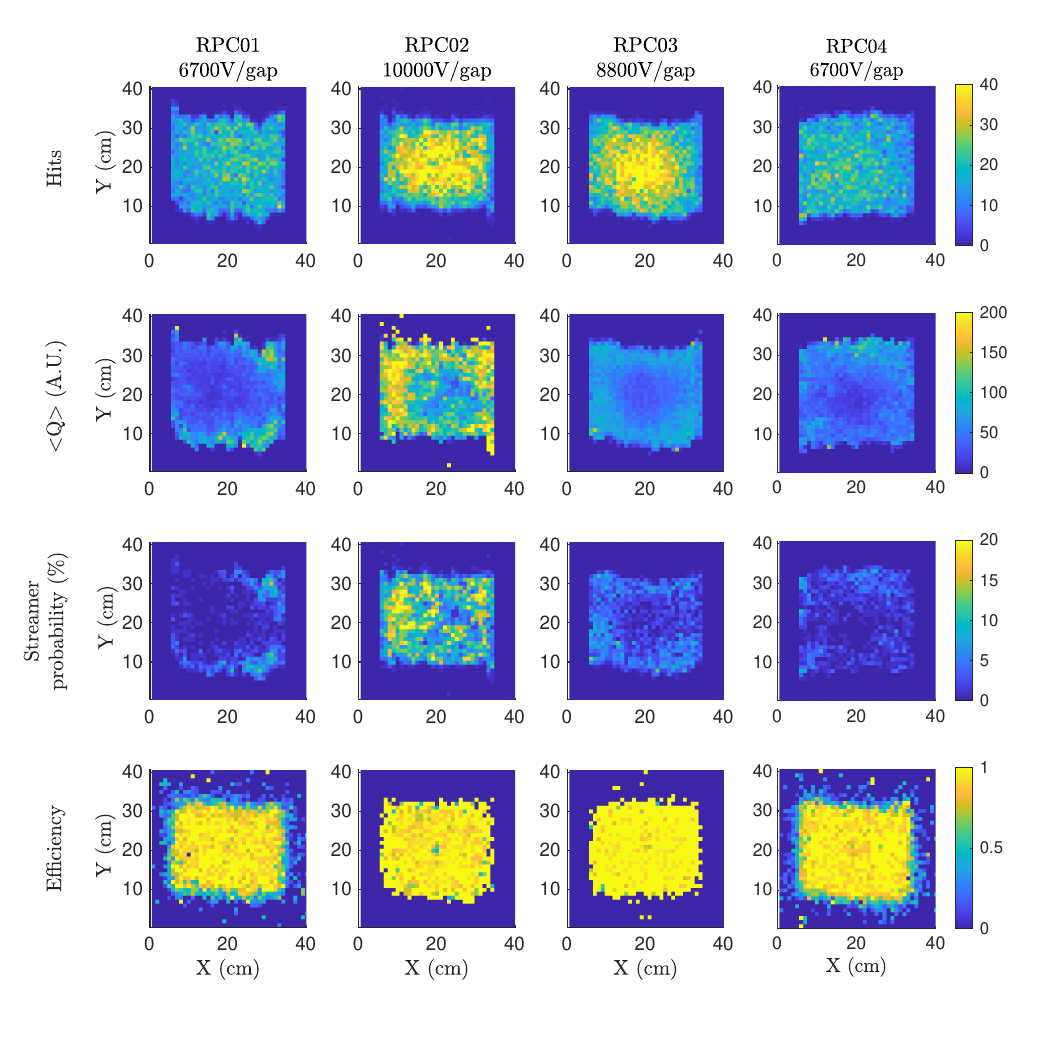}
	\caption{Detected hits, average charge, streamer probability and efficiency for the four planes at a voltages within the HV plateau.}
	\label{fig:XYscan}
\end{figure}

Figure \ref{fig:XYscan} shows the same variables as Figure \ref{fig:HVscan}, average charge, streamer probability and efficiency and in addition the number of detected muons (hits) as a function of the position within each of the planes. The value of the HV/gap is $6700$~V, $10000$~V, $8800$~V and $6700$~V for RPC01, RPC02, RPC03 and RPC04 respectively. Values well within the HV plateau. The maps are not perfectly square due to lack of accurate calibration of the strips time offsets. Anyway, they show homogeneous behavior with no obvious correlations with position. The efficiency maps show a stronger edge scatter in the RPC01 and RPC04 planes due to lower tacking precision, an effect that does not exist in the inner planes. It is also evident the larger values of the mean charge and streamer probability for RPC02 as already mentioned before.

\begin{figure}
	\centering 
	\includegraphics[width=\linewidth]{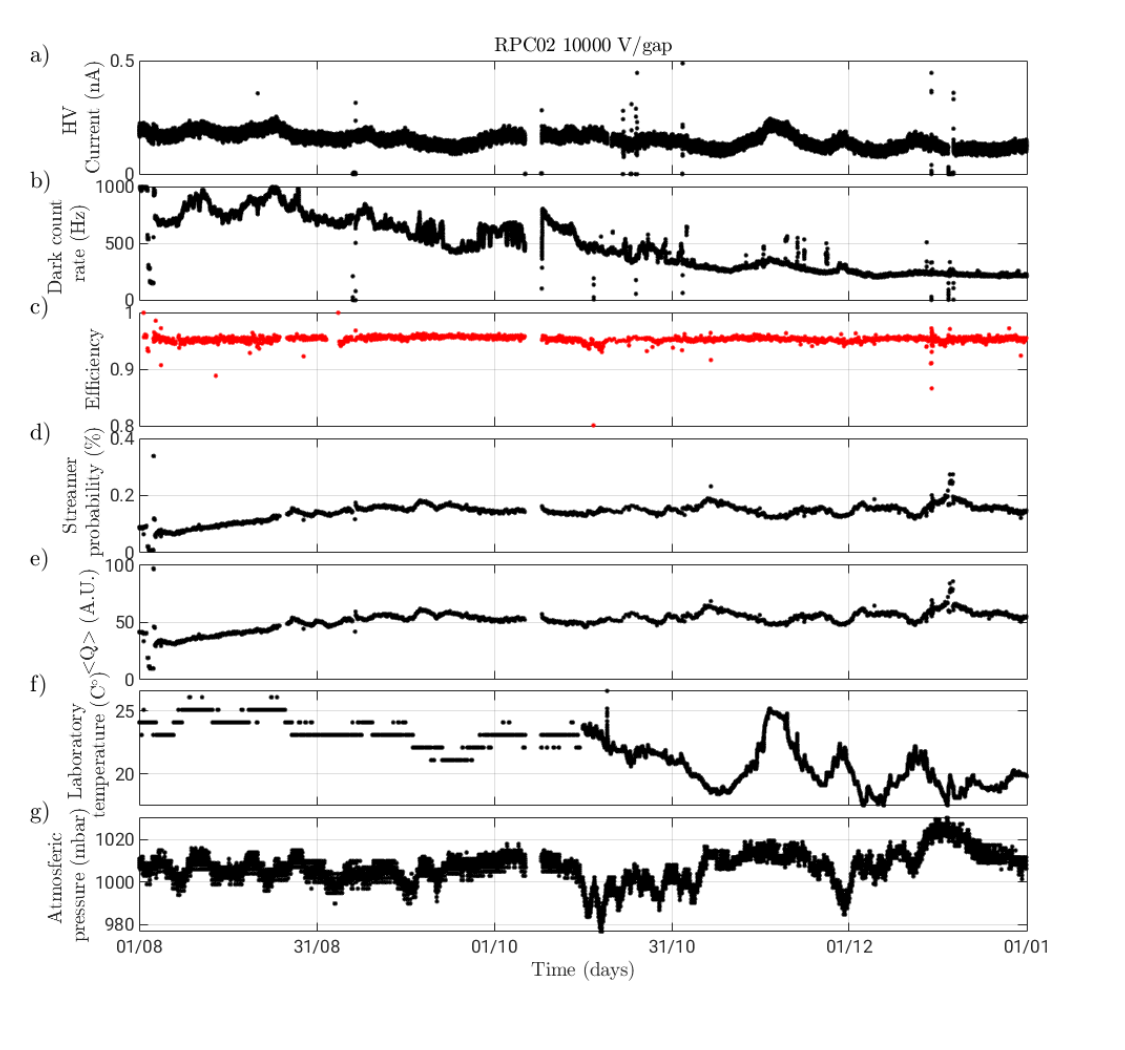}
	\caption{Relevant quantities that characterizing the stability of a sRPC plane (RPC02) for a period of five months. a) current at the HV power supply, b) dark count rate, c) efficiency, d) streamer probability, e) average charge, f) laboratory temperature and g) atmospheric pressure.}
	\label{fig:timescan}
\end{figure}

Figure \ref{fig:timescan} shows the same variables as Figure \ref{fig:HVscan}, average charge, streamer probability and efficiency, and in addition, HV power supply current and dark count rate (background) together with two environmental variables: laboratory temperature and atmospheric pressure, as a function of time for a period of five months. The plots correspond to RPC02 at a voltage of $10000$~V/gap. Other planes show similar behavior. The current at the HV power supply, figure \ref{fig:timescan}.a, is stable although modulated, basically, by the temperature,  figure \ref{fig:timescan}.f, due to the DC leakage currents, strongly dependent on it. The dark count rate (background), figure \ref{fig:timescan}.b, has decreased substantially (more than a factor 2) and stabilized probably due to the detector conditioning process. The efficiency, average charge and streamer probability, figures \ref{fig:timescan}.e, \ref{fig:timescan}.d and \ref{fig:timescan}.c, have remained stable throughout the period, except for initial transient of the last two at the beginning of the period without apparent continuity.

\section{Conclusions}
Four sRPC modules, a RPC requiring no continuous gas flow for its operation, of $300$~x$300$ mm$^2$ constituted by a multi-gap structure with $2$ gaps of either $1$~mm, $1.5$~mm and $2$~mm  has been operated for approximately five months without apparent degradation of the detector main characteristics and without the need for fresh gas supply.

The results shows a performance comparable to what could be expected from similar detectors operated in a continuous gas flow, efficiency higher than $95$\% and streamer probability below a few percent. The spatial dependence of efficiency, mean charge and streamer probability do not reveal significant structures. Similarly, these magnitudes do not seem to depend on the operating time (at least in the measured time interval), remaining stable over time.

These results seem to point to the possibility of the operation of multi-gap RPCs i9 for extended periods of time without fresh gas supply when exposed to the natural cosmic ray flux.

\section{Acknowledgments}
This work was supported by Funda\c{c}\~ao para a Ci\~encia e Tecnologia, Portugal in the framework of the project CERN/FIS-INS/0006/2021

\bibliographystyle{elsarticle-num-names}
\bibliography{paper-bibliography.bib}

\end{document}